# Enhanced mechanical performance and bioactivity in strontium / copper co-substituted diopside scaffolds

Shumin Pang [a], Dongwei Wu [b], Haotian Yang [a], Franz Kamutzki [a], Jens Kurreck [b], Aleksander Gurlo [a], Dorian A. H. Hanaor [a]*

[a] Technische Universität Berlin, Chair of Advanced Ceramic Materials, Straße des 17. Juni 135, 10623 Berlin, Germany

[b] Technische Universität Berlin, Chair of Applied Biochemistry, Gustav-Meyer-Allee 25, 13355 Berlin, Germany

* Corresponding author

## Abstract

Effective scaffolds for bone-tissue-engineering are those that combine adequate mechanical and chemical performance with osseointegrative, angiogenetic and anti-bacterial modes of bioactivity. To address these requirements via a combined approach, we additively manufactured square strut scaffolds by robocasting precipitation-derived strontium/copper co-substituted diopside. Microstructure, mechanical performance, bioactivity, biocompatibility, and antibacterial activity were examined. The results show that the presence of strontium and copper in the diopside lattice reduces the grain size and increases the density of the ceramics. The compressive strength, hardness, and fracture toughness of the diopside showed improvement, attributed to a finer microstructure and improved sintering. Scaffolds had excellent compressive strength with a high porosity (68-72%), which is attributed to the structure of the stacked-square-struts. All materials showed good in vitro bioactivity and favorable proliferation of osteogenic sarcoma cells, while strontium/copper co-doped materials exhibited the strongest anti-Escherichia coli activity. We show that across multiple indicators this system offers pathways towards high-performance bone substitutes.

**Keywords**: Bone-Tissue-Engineering; Diopside; Bioceramics; Bioactive materials; Antibacterial activity

## 1. Introduction

The field of bone-tissue-engineering has undergone significant development in the past two decades. New understandings of tissue regeneration and bioactive materials have paved the way for the application of implantable ceramic scaffolds into which bone-tissue may grow. Such tissue engineering can facilitate the treatment of large, hitherto untreatable bone defects arising from tumors, trauma and congenital defects. Polycrystalline silicate ceramics have been emerging in recent years as attractive biomaterials [1]. Although the solubility of bioceramics is generally lower than that of bioglasses, the diverse functionality that has been achieved in these materials, alongside their generally favorable mechanical properties motivates their use as bone-tissue-engineering materials [2, 3]. When considering the fracture toughness of bioceramics, diopside stands out from other ceramics with values in the range of $2.8 - 4$ MPa m$^{1/2}$ [4-6]. Our recent study on copper modified diopside was in agreement with this range of fracture toughness and showed some further toughening as the result of copper inclusion, possibly as the result of enhanced sintering behavior, which is typical of cation-substituted compounds [7]. Numerous studies have shown that copper ions and their release from bioceramics accelerate vascularization, bone tissue regeneration and hinder various types of bacterial infection [8-11]. However, the amount of copper included in bioceramics should not exceed a certain limit, otherwise, it will disrupt the intracellular balance and produce cytotoxicity. We have recently reported that copper substitution for magnesium in well-sintered diopside was cytocompatible up to 1 at.% relative to magnesium [7].

In numerous ceramic structures, strontium has been shown to readily substitute at calcium sites, unsurprisingly as strontium lies directly below calcium on the periodic table [12-14]. Substitutional strontium doping has been successfully explored in diverse systems as an avenue towards improving osseointegration, vascularisation, biodegradation, and fracture toughness of bioceramics [15, 16]. In bone tissue engineering, of particular value is the intrinsic tendency of bioavailable strontium to stimulate bone regeneration and osteogenetic differentiation, proposed to be mediated by the modulation of the Akt/mTOR signalling pathway, while hindering bone resorption through the inhibition of osteoclast activity [17, 18]. Reportedly, these modes of bioactivity may be further enhanced through a synergy of strontium and silicon in bioceramics [19]. Confirming these mechanisms, the inclusion of strontium in biomaterials of various forms has shown promising results towards the treatment of osteoporosis in a raft of *in vitro* and *in vivo* studies [20, 21].

Co-substitution, where multiple different cations are substituted at non-equivalent sites in a ceramic lattice, poses an effective strategy towards the extraction of increased levels of multi-mode bioactive performance in a given crystalline bioceramic phase, while avoiding potentially problematic effects that may arise from an excessively high level of a single dopant, such as cytotoxicity for the case of copper. By inducing lattice distortion and fluxing effects, co-substitution is a logical approach to achieving simultaneous improvements to biodegradability and sinterability of crystalline bioceramics. The rationale of the present study is therefore to implement the co-substitution of copper and strontium at the M1 (magnesium) and M2 (calcium) octahedral sites respectively in the clinopyroxene structure of diopside, with a view towards achieving a tough material with high bioactive performance. The similarity in ionic size of calcium/strontium and magnesium/copper is conducive to this substitution scheme and indeed in akermanite, full substitution of strontium at calcium sites and copper at magnesium sites is known to be possible [22].

Additive manufacturing (AM) technology allows the fabrication of scaffolds with tailored geometries and is increasingly being used in clinical treatments [23]. Robocasting is an AM technology based on the direct extrusion of slurry that allows the fabrication of ceramic green bodies in diverse and complex forms, which circumvents the limitations of conventional scaffold production techniques to accommodate complicated bone defects [24, 25]. Traditionally, a nozzle with a circular outlet is used to print scaffolds in woodpile structures or similar geometries, which leads to stress concentration and is not conducive to manufacturing a high-strength scaffold, whereas a nozzle with a square outlet can overcome this drawback. In this work, considering the mechanical strength and biocompatibility of polycrystalline diopside and inspired by the opportunities afforded by the inclusion of strontium and copper in bioceramics, we fabricated square strut diopside scaffolds co-doped with strontium and copper using a robocasting approach, and evaluated the mechanical properties, antibacterial activity and biocompatibility of the scaffolds. The results presented here show how this design rationale may facilitate multifunctional bioceramics for bone-tissue-engineering.

## 2. Materials and methods

### 2.1. Synthesis of powders

Diopside (here named DIO) powders were synthesized by a coprecipitation method [26]. Briefly, magnesium chloride hexahydrate ($MgCl_2 \cdot 6H_2O$, Carl-Roth, Germany, > 98%) and calcium chloride ($CaCl_2$, Carl-Roth, Germany, >98%) were dissolved in ethanol. Tetraethyl orthosilicate (($C_2H_5O)_4Si$, TEOS, VWR Chemicals) was subsequently added to the solution to achieve the atomic ratio of Ca: Mg: Si= 1: 1: 2, and the resulting solution was stirred for 2 h. Precipitation of the silicate was achieved by the addition of 25 wt.% aqueous ammonia (Merck) into the solution and the resulting suspension was stirred overnight. The precipitated product was centrifuged, washed, dried overnight at 90°C, and calcined at 900°C for 2 h. Similar steps were taken to synthesize strontium substituted diopside powders by replacing the stoichiometric amount of $CaCl_2$ with 2 at.% of strontium chloride ($SrCl_2$, Sigma-Aldrich, 98%) to produce $Ca_{0.98}Sr_{0.02}MgSi_2O_6$ (named Sr-DIO). To achieve co-doped diopside, $SrCl_2$ and copper chloride ($CuCl_2$, Sigma-Aldrich, 99%) were added at an atomic ratio of Ca: Sr: Mg: Cu: Si = 0.98: 0.02: 0.99: 0.01: 2, giving the composition of $Ca_{0.98}Sr_{0.02}Mg_{0.99}Cu_{0.01}Si_2O_6$ (named Sr/Cu-DIO). The synthesized powders were collected and sieved to a particle size smaller than 63 μm for subsequent use.

### 2.2. Design and production of the extrusion nozzle

DIO, Sr-DIO and Sr/Cu-DIO scaffolds were fabricated by robocasting using a customized extrusion nozzle to form a woodpile structure with square-cross-section members. Fig. 1 displays a schematic diagram of the process for manufacturing the scaffolds. As shown in Fig. 1, the extrusion nozzle was designed with a female Luer inlet and its outlet was designed as a square with a side length of 1.2 mm (a round outlet nozzle with the same cross-sectional area was also designed as a control). Then, a stereolithographic (SLA) 3D printer (Form 2, Formlabs) with clear photoreactive liquid resin (Clear Resin V4, Formlabs) was utilized to produce the nozzles. After printing, they were rinsed with isopropanol and post-processed with a UV curing apparatus to ensure sufficient curing.

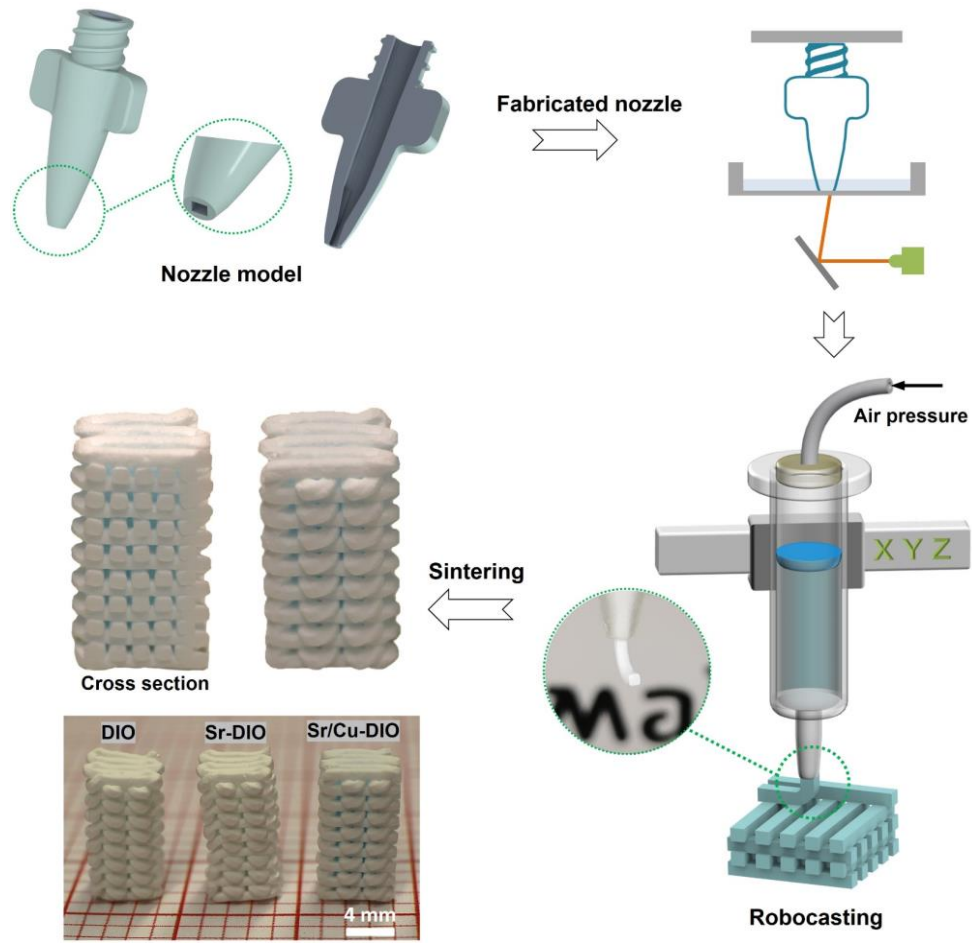

Fig. 1. Schematic illustration of the preparation procedure of square strut scaffolds.

## 2.3. Preparation of slurry and robocasting of scaffolds

To prepare the printable slurry, 10 g quantities of the synthesized powders were mixed with 0.4 g sodium alginate (low viscosity, Alfa Aesar) and 8.0 g 20 wt.% Pluronic F-127 solution (Sigma-Aldrich). The prepared slurry was transferred into a syringe (Vieweg GmbH, German), to which our customized nozzle was coupled. The syringe was tapped vigorously to remove bubbles and fixed onto a robotic deposition 3D printer, which was modified and re-equipped with an air-pressure controller based on an Ultimaker 2 go printer (Ultimaker BV, Netherlands). The green bodies were transported into a covered container with a small opening after printing for slow drying for 4 days, which is crucial to prevent cracking and ensure homogeneous drying. Finally, the dried green bodies were sintered in the air at 1250° for 3 h.

## 2.4. Characterization

X-ray diffraction (XRD) patterns of sintered scaffolds were collected by an X-ray diffractometer (D8 advance, Bruker AXS, USA) at 35 kV and 40 mA using CuKα radiation in 3 s/0.02° steps. The functional groups of the sintered powders were examined using a Fourier transform infrared spectrometer (FTIR, Bruker Equinox 55, Germany) in attenuated total reflection (ATR) mode. By employing monochromatic Al K Alpha at 1 μA and 2 kV during X-ray photoelectron spectroscopy (XPS) on an Escalab 250250 (Thermo Fisher Scientific, USA), the chemical state and elemental composition of the samples were investigated. The calibrated C1s for the XPS spectra is 284.8 eV. Inductively coupled plasma optical emission spectroscopy (ICP-OES) was used to confirm the quantities of strontium and copper in Sr-DIO and Sr/Cu-DIO samples in an ICP Ultima2 (Horiba, Japan). To achieve this, samples were digested in an autoclave at 220 °C for 24 h with the addition of $H_2SO_4$, $H_3PO_4$, and HF in an aqueous solution. The microstructure of the sintered scaffolds was studied using scanning electron microscopy (SEM, LEO Gemini 1530, Germany) with acceleration voltage varying from 3 to 10 kV. An LS 13320 laser diffraction particle size analyzer (Beckman Coulter, USA) was used to analyze the particle size in an ultrasonically dispersed aqueous solution. The apparent density and apparent porosity of the scaffolds were estimated using the Archimedes method based on the following equations [27]:

$$\text{Apparent density} = \frac{W_d}{W_w - W_s} \qquad (1)$$

$$\text{Apparent porosity (\%)} = \frac{W_w - W_d}{W_w - W_s} \times 100 \qquad (2)$$

where $W_w$ represents the weight of the scaffold with water, $W_d$ is the dry weight of the scaffold, and $W_s$ represents the wet weight of the scaffolds suspended in water. Six replicates of the measurements were performed.

The Vickers hardness ($H_V$) and fracture toughness ($K_{IC}$) values of the scaffolds were measured on polished surfaces using a microhardness indenter (Zwick, 3212001/00, Germany) in accordance with ISO 21618 standard, by applying a 5 kg load with a dwell time of 15 s. Values of $H_V$ and $K_{IC}$ were calculated as:

$$H_V = 0.001854 \frac{F}{(2a)^2} \qquad (3)$$

$$K_{IC} = 0.000978 \left(\frac{E}{H_V}\right)^{0.4} \left(\frac{F}{c^{1.5}}\right) \qquad (4)$$

Where F is the applied force (N), a is the mean half-length diagonal value (mm), E is Young's modulus, which was evaluated by a TriboIndenter (Hysitron TI950, Bruker Corporation, USA) equipped with a standard Berkovich diamond indenter tip, c is the mean half crack length (mm). Ten indentations for each group were used to obtain the average values along with their standard deviations.

The compressive strengths of bulk materials and scaffolds were determined on the basis of sintered pellets with 8 mm diameter and additively manufactured square- and round-strut scaffolds ($5 \times 5 \times 10$ mm³). Compression of specimens was conducted between flat platens at a crosshead speed of 0.5 mm/min using a mechanical testing machine equipped with a 10 kN load cell (Zwick/Roell, Germany). Measurements were conducted in eight replicates.

## 2.5. *In vitro* bioactivity evaluation

To assess the osseointegrative bioactivity of the scaffolds, the scaffolds were immersed in simulated body fluid (SBF) at 37°C for 28 days, as proposed by Kokubo et al. [28]. During the soaking period, the SBF solution was refreshed every 3 days. After 28 days of soaking, the scaffolds were dried at 60°C for 24 h and evaluated by SEM (Zeiss DSM 982 Gemini, Germany) and XRD.

## 2.6. *In vitro* cell studies

The influence of Sr/Cu-DIO scaffolds on cell behavior was studied using human osteogenic sarcoma cells (Saos-2 cells, Charité, Department of Periodontology, Oral medicine and Oral surgery). The Saos-2 cells were cultured in an incubator with a humidified atmosphere (37°C, 5% $CO_2$). For cell culture, a complete medium was used and replaced every other day. Specifically, Dulbecco's Modified Eagle medium (DMEM, High Glucose, Biowest) with 15% fetal bovine serum (FBS) and 2 mM L-Glutamine was used for Saos-2 cells. Scaffold extracts were prepared by immersing 2 g scaffold in 10 mL serum-free medium and incubating at 37°C for 24 h. The supernatant was then collected and subsequently sterilized through a filter (Millipore, 0.22 mm), followed by supplementation of 10% FBS.

## Cell viability assay

2,3-bis (2-methoxy-4-nitro-5-sulfophenyl)-2H-tetrazolium-5-carboxanilide sodium salt (XTT sodium salt, Alfa Aesar) was used to assess cell viability after exposure to scaffold extracts, as described by Wu et al. [29]. Briefly, after seeding the cells in 96-well culture plates, the supernatant in the experimental well was changed for 100 µL scaffold extract, and XTT solution with phenazine methosulfate (PMS, AppliChem, Germany) was added to each experimental well after 1, 3, and 7 days of incubation. After 4 h of incubation, the absorbance of the supernatant at 450 nm was determined using a microplate reader (Tecan Austria GmbH). As a control, fresh medium was tested under the same conditions. And fresh medium with XTT solution was set as background. The test was conducted in six replicates.

To assess cell viability on the scaffolds, Saos-2 cells were stained using a live/dead staining kit (L3224, Thermo Fisher Scientific). Briefly, after seeding cells on the scaffold for 48 h, the scaffolds were rinsed with phosphate buffered saline (PBS), followed by the addition of 2 µM ethidium homodimer-1 and 2 µM calcein AM to the experimental wells and incubation in an incubator for 15 min. Fluorescence microscope (Observer Z1, Zeiss) was then used to observe fluorescence in three repetitions. Green fluorescence represents living cells, while red fluorescence indicates dead cells.

**Immunofluorescence staining**

Saos-2 cells were seeded onto the scaffolds and cultured for 48 h prior to immunofluorescence staining of Actin. Concisely, the scaffolds were fixed with Histofix Eco solution (ROTI® Histofix ECO Plus, Roth) for 15 min and washed with PBS for two times. The samples were then permeabilized with 0.1% Triton X-100 for 15 min and washed with PBS twice, as well as blocked with 1% bovine serum albumin for 30 min at room temperature. Subsequently, the filamentous actin (F-actin) was stained with rhodamine phalloidin (Invitrogen™, 1:400) for 2 h, followed by nuclear labelling with 4′,6-diamidino-2-phenylindole (DAPI, Sigma-Aldrich, 1 mg/mL) for 30 min. After each staining, the scaffolds were rinsed with PBS so as to reduce the imaging background. Finally, the cells were observed under a fluorescence microscope and the cell spreading area on the scaffolds was measured using the *Image J* software.

## 2.7. Antibacterial activity

We investigated the antibacterial activity of Sr/Cu-DIO scaffolds against Escherichia coli (E. coli, ATCC 25922) by the liquid medium microdilution technique and the agar diffusion method. Regarding the liquid medium microdilution method, 0.3 g scaffold was soaked for 24 h with bacterial suspensions ($1 \times 10^4$ CFU/mL). A UV-Vis spectrophotometer was used to measure the absorbance values of all the tested solutions at 600 nm. The untreated bacterial solution was set as a control. Six parallel samples were used for each test. The following equation was used to calculate the percentage of bacterial inhibition (3):

$$\text{Inhibition (\%)} = \frac{I_C - I_S}{I_C} \times 100 \qquad (3)$$

where $I_S$ and $I_C$ are the absorbances of the scaffold bacterial suspensions and the control, respectively.

The antibacterial activity of the scaffolds was further assessed using the agar diffusion method. To this end, a 100 µL E. coli suspension ($1 \times 10^8$ CFU/mL) was seeded on a Luria-Bertani (LB) agar medium plate. By soaking 0.3 g pieces of scaffold in LB medium for 24 h at 37°C, scaffold extracts were obtained. Then, sterilized 15 mm-diameter filter paper discs were immersed in this LB medium. As a control, the filter paper was immersed in sterile deionized water. Subsequently, these filter paper discs were put on the surface of the E. coli seeded agar medium and incubated at 37°C for 24 h. Finally, *ImageJ* software was used to analyze photographs and measure the inhibition zone of bacterial growth (the diameter of the inhibition zone included the diameter of the filter paper). Each test was performed with three parallel samples.

## 2.8. Statistical analysis

In this work, all data were expressed as a mean value with a standard deviation (SD). The differences between experimental groups were analyzed using the analysis of variance. *$p < 0.05$ and **$p < 0.01$ denote that the results are significantly different.

## 3. Results

### 3.1. Structural characterization of scaffolds

Fig. 2a shows the XRD patterns of sintered scaffolds. It can be observed that DIO, Sr-DIO, and Sr/Cu-DIO scaffolds previously.

consist of single-phase diopside (PDF#72-1497), with no discernible presence of other crystalline phases. In order to show the effect of strontium and copper substitution on the diopside lattice, the enlarged patterns of the sharpest peak between $29.5°-30.1°$ are displayed in Fig. 2b. In comparison to the DIO material, diffraction peaks from Sr-DIO and Sr/Cu-DIO samples are shifted to slightly lower $2\theta$ values indicating lattice expansion, which is due to the larger ionic radius of $Sr^{2+}$ (1.26 Å, CN=8) relative to $Ca^{2+}$ (1.12 Å, CN=8), and the ionic radius values are taken from R. D. Shannon [30]. Due to the similar radii of $Cu^{2+}$ (0.73 Å, CN=6) and $Mg^{2+}$ (0.72 Å, CN=6) in octahedral coordination, the substitution of copper at magnesium sites is not anticipated to cause considerable lattice distortion at low doping levels as confirmed in the $2\theta$ range of $29.5° - 30.1°$. (c) FTIR spectra of DIO, Sr-DIO, and Sr/Cu-DIO samples in the range of $1800 - 400 \ cm^{-1}$.

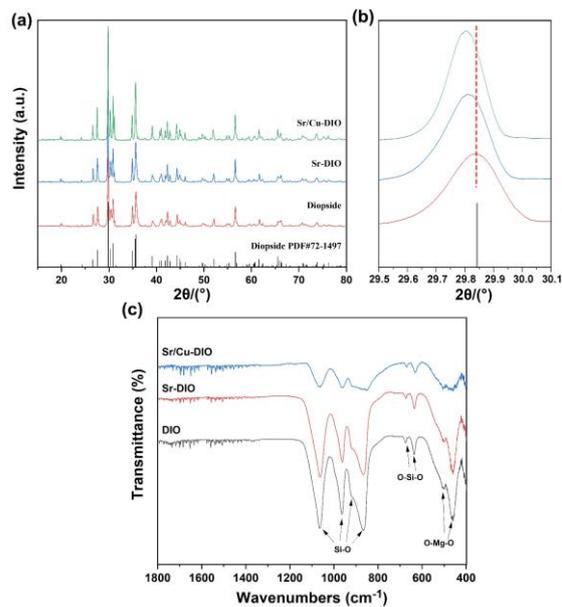

Fig. 2. (a) XRD patterns of DIO, Sr-DIO, and Sr/Cu-DIO samples sintered at 1250°C, (b) peak shift shown

The FTIR spectra of the sintered samples are shown in Fig. 2c to confirm the XRD analysis and verify the insertion of strontium and copper into the structure. In DIO, the peaks at 462 and 502 $cm^{-1}$ were attributed to the bending vibrations of O-Mg-O. The bending vibration of O-Si-O was responsible for the peaks at 635 and 674 $cm^{-1}$. Additionally, two other peaks at 866 and 918 $cm^{-1}$ corresponded to the stretching mode of Si-O. The symmetric stretching of Si-O was also attributed to the peaks at around 962 and 1065 $cm^{-1}$. The vibrations of the functional groups of the diopside are in good agreement with all of the measured FTIR peaks [31], consistent with the XRD analysis. The other major parts of the Si-O vibrations are not shifted, except for the peak shift of the Si-O stretching mode to 858 cm$^{-1}$ for the Sr/Cu-DIO sample, and the peak shift of the symmetric stretching of Si-O to 1062 cm$^{-1}$ and 1060 cm$^{-1}$ for the Sr-DIO and Sr/Cu-DIO samples, respectively. The observed are consistent with the presence of strontium or strontium/copper in the diopside structure.

change in wavenumbers of the Si-O group is due to the fact that Sr$^{2+}$ has a larger ionic radius than Ca$^{2+}$ and therefore partially reduces the symmetry of the Si-O bond when calcium is partially substituted by strontium in the diopside lattice, leading to a change in vibrational frequency [32]. For Sr/Cu-DIO, due to the partial replacement of magnesium by copper, the O-Mg-O displayed a decreased intensity[33]. These observations

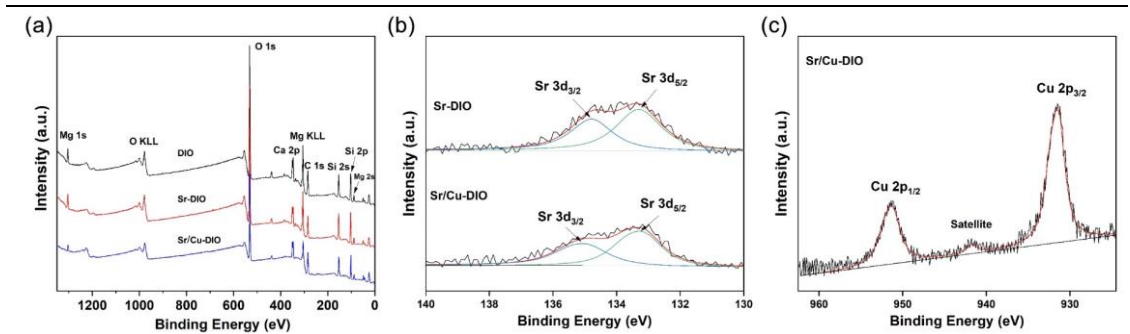

Fig. 3. (a) XPS survey spectra of DIO, Sr-DIO, and Sr/Cu-DIO. Fitting spectra of (b) Sr 3d, (c) Cu 2p.

The surface elemental composition as well as the chemical state of the sample were evaluated by XPS. Fig. 3a shows typical XPS survey scans of DIO, Sr-DIO, and Sr/Cu-DIO, all of which contained the expected components of diopside, Ca, Mg, Si, and O. In Fig. 3b, the characteristic peak of Sr 3d was detected in both Sr-DIO and Sr/Cu-DIO samples, and the Sr 3d spectrum was deconvoluted into two double peaks, 3d$_{3/2}$ at 133.1 eV and 3d$_{5/2}$ at 134.8 eV, indicating the presence of Sr$^{2+}$. Interestingly, strontium doping slightly increases the binding energy of Ca 2p from 346.7 eV for DIO to 347.5 eV (2Sr-DIO) and 347.4 eV (2Sr/Cu-DIO), respectively, due to the entry of strontium into the diopside lattice and the substitution of calcium sites, which is consistent with previous studies [34, 35]. In addition, the Cu 2p energy spectrum was detected in 2Sr/Cu-DIO, as shown in Fig. 3c, with two main peaks at 951.4 eV and 931.5 eV, corresponding to Cu 2p$_{1/2}$ and Cu 2p$_{3/2}$, respectively. The characteristic satellite peak of Cu$^{2+}$ is located at 941.8 eV [36]. The Cu 2p spectrum is fitted to only one Gauss-Lorentz peak, indicating that the copper in the 2Sr/Cu-DIO sample is mainly in the form of Cu$^{2+}$ [37]. To confirm this observation, elemental analysis was performed by ICP-OES. Results indicated a strontium content in the Sr-DIO sample of 1.67 at.%, while the strontium content in the Sr/Cu-DIO sample is 1.66 at.%, and the copper content is 0.62 at.%, with theoretical values of 2 at.% and 1 at.%, respectively. The ICP-OES results show that strontium and copper are successfully incorporated into diopside, and while the experimental values are smaller than the theoretical values, this is likely caused by incomplete solubility of silicates during the test, a known problem for such materials.

Fig. 4a-c presents SEM micrographs of DIO, Sr-DIO, and Sr/Cu-DIO scaffolds. The morphology of diopside was not significantly changed by the substitution of strontium and copper, and the sintered scaffolds showed uniform microstructure with irregularly shaped grains. In comparison with unmodified diopside, a slight decrease in grain size was observed in the Sr-DIO and Sr/Cu-DIO materials on the basis of SEM images, which is consistent with the laser diffraction particle size analyzer result in Fig. 4d, while the high median size measured by particle size analyzer is most likely due to particle agglomeration during measurement [38]. The SEM results indicate that the strontium and copper substitution into the diopside structure limits the grain growth that occurs during sintering. Indeed, such ion substitution is known to inhibit grain growth by pinning grain boundaries in ceramic materials, thereby improving mechanical strength [39-41]. Furthermore, the sintered Sr/Cu-DIO exhibited a denser structure, an expected result of the greater fluxing which occurs through the co-substitution of different ions.

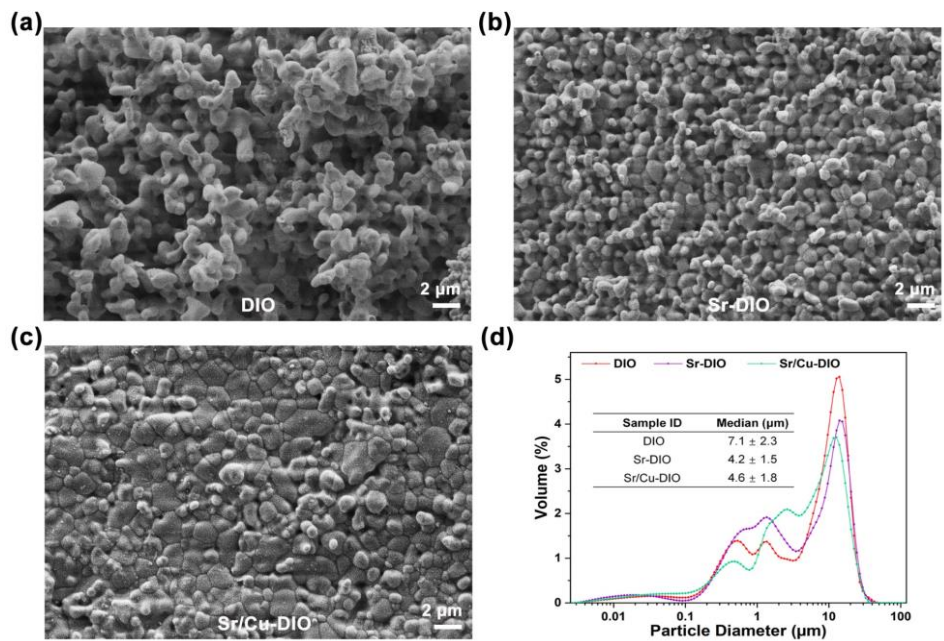

Fig. 4. Microstructure of sintered scaffolds. SEM Micrographs of (a) DIO, (b) Sr-DIO, (c) Sr/Cu-DIO; (d) Particle size distribution of sintered powders (n = 3).

## 3.2. Mechanical performance

Table 1 The properties of cancellous bone and sintered scaffolds.

| Sample | Apparent porosity (%) | Apparent density (g/cm³) | Vickers Hardness (GPa) | Fracture toughness (MPa·m^{1/2}) |
|---|---|---|---|---|
| Cancellous Bone [42, 43] | 30 - 90 | $0.06 - 0.22$ | $0.4 - 0.6$ | $0.1 - 0.8$ |
| DIO | $72.2 \pm 1.6$ | $1.29 \pm 0.05$ | $10.51 \pm 1.54$ | $3.01 \pm 0.82$ |
| Sr-DIO | $71.1 \pm 1.1$ | $1.30 \pm 0.06$ | $11.07 \pm 1.35$ | $3.07 \pm 0.94$ |
| Sr/Cu-DIO | $68.4 \pm 1.8$ | $1.32 \pm 0.05$ | $11.93 \pm 0.58$ | $3.23 \pm 0.31$ |

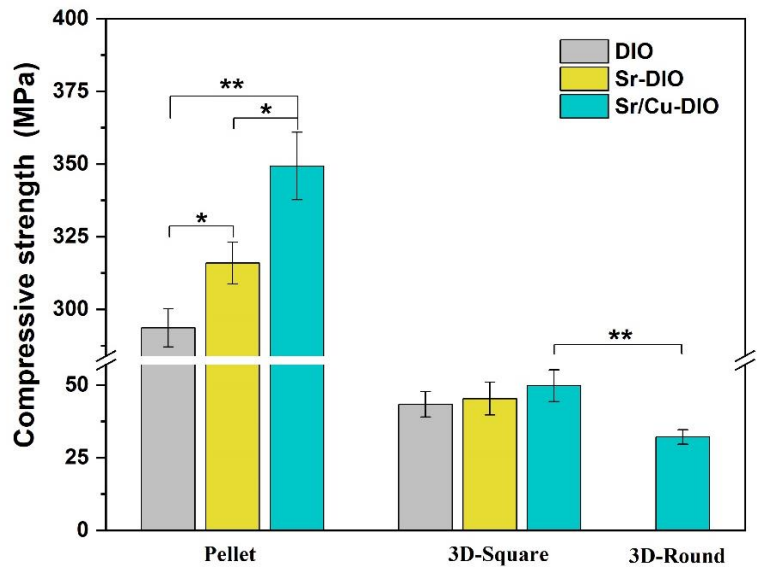

Fig. 5. The compressive strength of different types of materials after sintering: pellet, 3D printed square strut scaffold and 3D printed round strut scaffold (n = 8, *P < 0.05 and **P < 0.01).

The apparent porosity, density, Vickers hardness, and fracture toughness of DIO, Sr-DIO, and Sr/Cu-DIO scaffolds after sintering are shown in

Table **1**. The results show that the porosity of DIO, Sr-DIO, and Sr/Cu-DIO scaffolds are 72.2 ± 1.6, 71.1 ± 1.1, and 68.4 ± 1.8%, respectively, all of which would be adequate to induce effective cell ingrowth within the scaffold. As seen in microstructural analysis by SEM (Fig. 4), the substitution of strontium or the co-substitution of strontium and copper both increased the apparent density of the materials, a result of the aforementioned fluxing effect that cation substitution tends to have. The intrinsic mechanical properties of biomaterials play an important role in their processing and *in vivo* performance. The improved mechanical strength arising from the co-addition of strontium and copper ions can be related to their improved intrinsic fracture toughness, with a value of 3.23 ± 0.31 MPa m$^{1/2}$ being attained. As will be discussed, grain boundaries have a significant influence on crack resistance, therefore the rise in H$_V$ and

K$_{IC}$ values of Sr/Cu-DIO may be related to its dense structure and small grain size [44]. For the square strut woodpile structure scaffolds produced here, as shown in Fig. 5, the compressive strength improved with strontium and copper substitution, with values for DIO, Sr-DIO, and Sr/Cu-DIO scaffolds being 43.42 ± 4.36, 45.34 ± 5.65, and 49.75 ± 5.48 MPa, respectively, while the compressive strength of the round strut Sr/Cu-DIO scaffold is significantly lower. The extrinsic, structure-driven strength of bone scaffolds can be engineered through the design of fabrication processes. The use of square struts in the scaffolds produced here reduces the stress concentration that generally occurs in the more common robocasting of round strut profiles and allowed the attainment of higher levels of compressive strength [45]. In addition, in order to characterize the effect of strontium and copper inclusion on the compressive strength of bulk diopside materials, compressive tests were conducted on pellets, with results confirming the significant increase in strength by strontium doping or strontium/copper co-doping (Fig. 5).

### 3.3. *In vitro* bioactivity

The osseointegrative bioactivity of DIO, Sr-DIO, and Sr/Cu-DIO scaffolds was assessed by their ability to form precipitated apatite layers during immersion in SBF. Fig. 6 shows the SEM micrographs of the scaffolds after 28 days of immersion in SBF. Compared to the morphology before soaking (Fig. 4), the surfaces of the DIO, Sr-DIO, and Sr/Cu-DIO scaffolds were significantly altered, exhibiting extensive coverage by flake-like apatite precipitates, characteristic of bioactive ceramics. Further XRD measurements of the immersed scaffold showed peaks at 21.7°, 22.7°, 26.2°, and 31.8°, which are attributed to the (2 0 0), (1 1 1), (0 0 2), and (2 1 1) planes of hydroxyapatite (PDF#74-0565), confirming this and indicating high levels of bioactivity of the scaffolds.

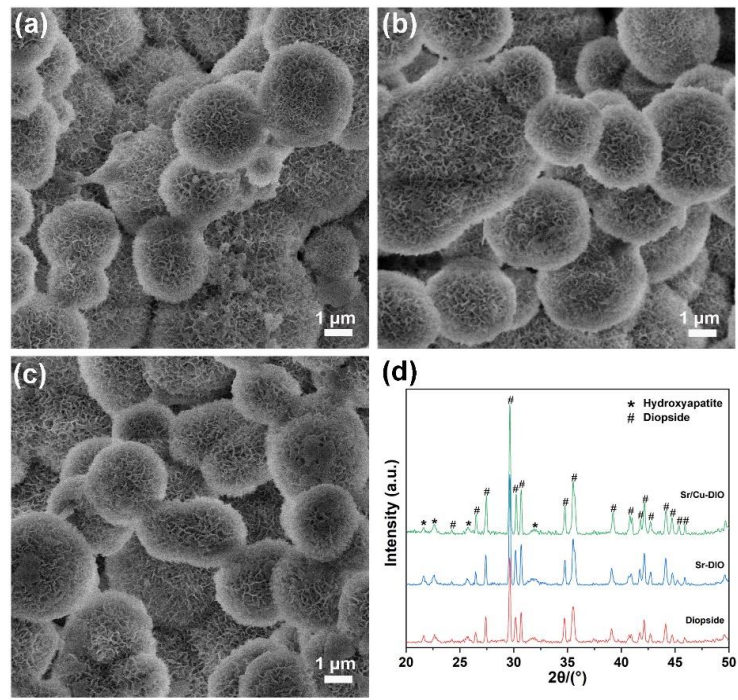

Fig. 6. SEM micrographs of (a) DIO, (b) Sr-DIO, and (c) Sr/Cu-DIO scaffold after 28 days of immersion in SBF. (d) And the XRD patterns of scaffolds after immersion.

### 3.4. Cytocompatibility

Fig. 7 shows the effect of ion extracts of DIO, Sr-DIO, and Sr/Cu-DIO scaffolds on the viability of the Saos-2 cells after 1, 3, and 7 days of incubation. All experimental groups showed higher cell viability than the control group at each set time point, with the highest observed on Sr-DIO, demonstrating the good cytocompatibility of this material. Interestingly, diopside co-doped with strontium and copper showed slightly lower cell viability than Sr-DIO, but higher than that of pure diopside.

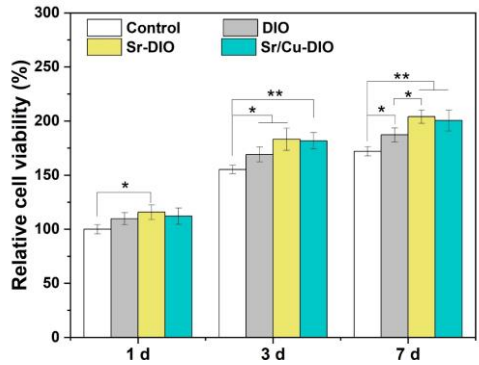

Fig. 7. *In vitro* cell viability after exposure to scaffold extracts for 1, 3, and 7 days, as determined by XTT assay (n = 6, *P < 0.05 and **P < 0.01).

We further observed the viability of Saos-2 cells on DIO, Sr-DIO, and Sr/Cu-DIO scaffolds by fluorescence microscopy. As indicated by the images shown in Fig. 8a-c, Saos-2 cells adhered well to Sr-DIO and Sr/Cu-DIO scaffolds, and the ratio of live/dead cells on the scaffolds shown in Fig. 8d is consistent with the results of the cell viability test. In addition, the morphology of Saos-2 cells on DIO, Sr-DIO, and Sr/Cu-DIO scaffolds was examined by immunofluorescence staining. The fluorescence micrographs clearly showed that the cytoskeletal structure of Saos-2 cells spread well on the scaffold after 48 h of culture (Fig. 8e-g). It can be seen that F-actin was observed in all scaffolds, indicating that the incorporation of strontium or strontium/copper into the diopside scaffold had no negative effect on the adhesion of Saos-2 cells. Further quantitative analysis showed that the spreading area of Saos-2 cells was greater for Sr-DIO and Sr/Cu-DIO scaffolds compared to the one for DIO scaffold, with Sr-DIO being the most significant (Fig. 8h). This finding suggests that Sr-DIO and Sr/Cu-DIO scaffolds have good *in vitro* biocompatibility, indicating their high level of appropriateness as bone scaffolds, enabling the in-growth of newly formed bone and blood vessels.

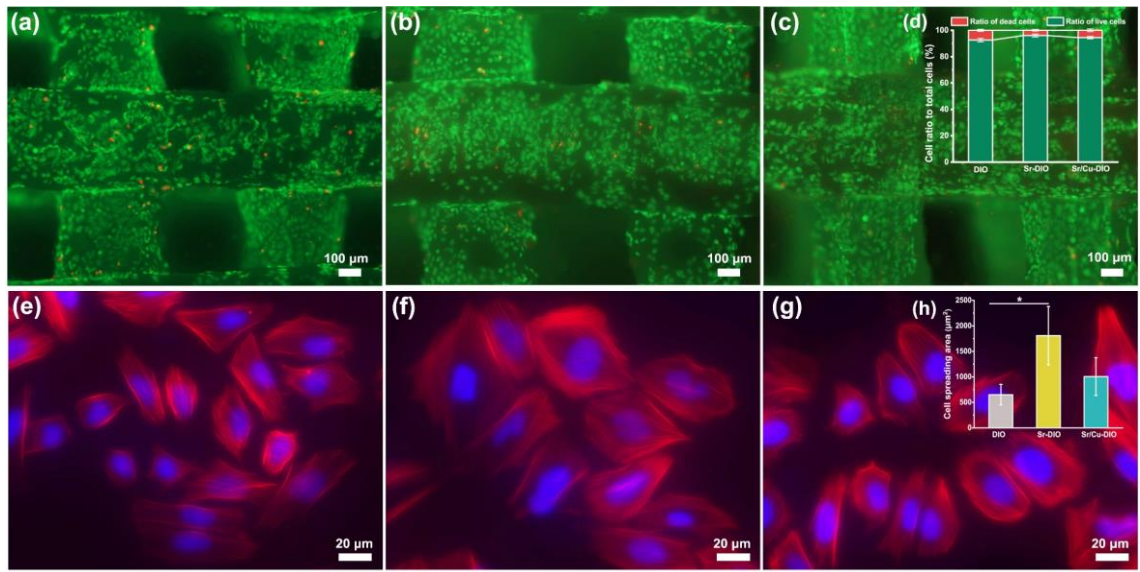

Fig. 8. Fluorescence images of live/dead staining of Saos-2 cells cultured on the (a) DIO, (b) Sr-DIO, and (c) Sr/Cu-DIO scaffold surface for 48 h. Green is used to indicate live cells, whereas red indicates dead cells. (d) Quantitative data of live/dead staining assay (n=3). Immunofluorescence staining of Saos-2 cells cultured on the scaffold for 48 h, (e) DIO, (f) Sr-DIO, and (g) Sr/Cu-DIO scaffold. F-actin was stained in red, and the nucleus was stained in blue. (h) Quantitative analysis of the cell spreading area on different scaffolds (n=3, *P < 0.05).

**3.5.**

**Antibacterial activity**

After 24 h of incubation, the antibacterial activity of DIO, Sr-DIO, and Sr/Cu-DIO scaffolds against E. coli was assessed. The antibacterial inhibition rate of DIO was around 42%, as indicated in Fig. 9a. Compared to DIO, the Sr-DIO scaffold showed stronger antibacterial activity, with the highest antibacterial activity observed on Sr/Cu-DIO scaffold. This result is further supported by Fig. 9b, which displays the results of antibacterial activity evaluated by the agar diffusion method and the quantitative results of the inhibition zone are presented in Fig. 9c. This also indicates that the native antibacterial activity of DIO was significantly enhanced by the inclusion of strontium and copper.

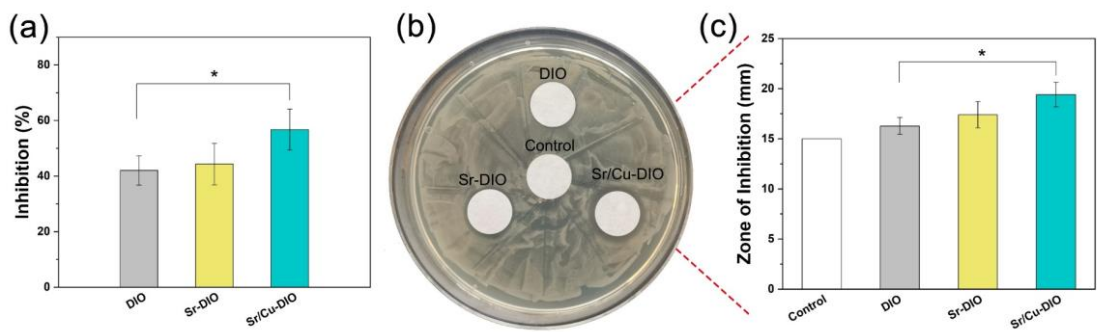

Fig. 9. (a) The antibacterial rate measured by the liquid medium microdilution method (n=5). (b) Representative images of E. coli on the filter paper containing scaffold extracts after 24 h of incubation on agar (n=3), and (c) the quantitative results of the inhibition zone evaluated by the agar diffusion method (*P < 0.05).

## 4. Discussion

Of the various bioactive silicates, those containing magnesium often draw attention due to their good mechanical properties, biodegradability, and biocompatibility. In this study, we investigated the fabrication and functionality of 3D printed bioceramic scaffolds based on strontium and copper doped diopside. Strontium and copper are readily incorporated into the single-phase diopside structure, substituting for calcium at the M2 site and for magnesium at the M1 site of the pyroxene structure. The microstructure of substituted and co-substituted materials was moderately altered with these having a higher density, fewer voids, and smaller grains. Strontium/copper co-doped samples clearly showed greater densification than the single-site doped material, an expected consequence of the fluxing effect of mixed cation occupancy. Fracture toughness and Vickers hardness followed the same trend as densification in the studied materials, with the co-doped materials showing a moderate improvement to unmodified diopside. Based on commonly observed trends in the fracture of polycrystalline ceramics, it can be postulated that enhanced sintering and finer microstructure are what contributed here to the higher fracture toughness in these materials, which followed the same trend as density and microstructure fineness [14, 46-48].

The fracture toughness values measured in materials fabricated here are consistent with previous reports showing a relatively high fracture toughness for diopside ceramics [4-7]. For co-substituted materials here, fracture toughness in terms of $K_{IC}$ was improved and reached approximately 3.23 MPa·m$^{1/2}$. These values are among the highest reported for a non-fibre-reinforced bioactive ceramic, and compare favorably even to some of the toughest bioglass ceramics [49, 50]. For bioactive ceramics, higher values of $K_{IC}$ enable the fabrication of higher surface area structures that are conducive to accelerated resorption and osseointegration while maintaining the necessary application-specific mechanical performance. In this study, scaffolds were produced by robocasting of a woodpile structure with square cross-sectional member profiles as opposed to the more common round profile struts used in such AM methods. The production of scaffolds having square profile members results in lower stress concentration at strut intersections and further contributes to a higher level of compressive strength in such scaffolds compared with scaffolds produced with round-profile members (Fig. 5). A biological scaffold material's porosity threshold is around 60%, below which it does not offer enough space for cell growth [51]. In this work, scaffolds with porosity greater than 68% showed higher compressive strength and hardness than cancellous bone [42].

An essential requirement for materials in bone-tissue engineering scaffolds is the ability to form a bioactive apatite layer on the surface, which facilitates binding to growing tissue, and effective bone regeneration. This is known as osseointegrative bioactivity. In this work, after 28 days of immersion in SBF, hydroxyapatite layers were clearly observed on the surface of all scaffolds, indicating the ability of scaffolds from these materials to maintain a strong interface with growing human bones and effectuate the healing. For osteoconductive and angiogenic bioactivity, cell adhesion and growth are important characteristics, with cell viability being mainly influenced by the ionic dissociation products of the biomaterial. Certain concentrations of strontium and copper doping have been shown to be beneficial for osteogenic differentiation and bone regeneration *in vivo* and *in vitro* [21, 52, 53]. Yin et al. produced borate-based glass scaffolds doped with different concentrations of strontium, confirming that the presence of strontium suppressed the rapid release of borate, significantly improving cell proliferation and reducing cytotoxicity [54]. Zhao et al. developed a strontium-doped calcium phosphate bioceramic that promotes local bone regeneration and osseointegration, which is a safe bone substitute for the treatment of bone defects [34]. In this work, 2 at.% strontium-doped diopside clearly enhanced cell viability and proliferation, which is consistent with observations in previous studies of strontium-doped bioceramics. However, when strontium and copper were co-doped into diopside (Sr/Cu-DIO), the cell viability was lower than that of Sr-DIO. This may be due to the distorted lattice which allows a more rapid release of copper ions into the medium, thus increasing solubility. In our previous study, 1 at.% copper doped diopside was demonstrated to have enhanced cell viability [7], and magnesium, calcium, silicon and strontium ions have been shown to be beneficial elements for bone regeneration [16, 55]. However, above a certain level, the high concentration of ionic dissociation products leads to high pH values, which can be detrimental to cells [14]. Nevertheless, the cell viability of the Sr/Cu-DIO scaffold is better than that of pure diopside and cells exposed to this material spread over a larger area, demonstrating that the co-doping of strontium and copper had a favorable effect on the biocompatibility of diopside.

A further important mode of bioactivity in materials used for bone repair is antibacterial activity, which serves to inhibit infection in the region of the bone defect. Choudhary et al. showed that diopside nanopowder had a native inhibitory effect on gram-positive and negative bacteria due to the release of calcium and magnesium ions from the diopside which increased the pH in the bacterial medium, thereby inhibiting the growth of bacteria [56]. Recent studies have shown that strontium inhibits the growth of E. coli, and the inhibition of Porphyromonas gingivalis and Staphylococcus aureus has also been reported [57-59]. In our work, we can postulate that the 2 at.% strontium doping (Sr-DIO) improved the inhibition of E. coli by DIO through the enhanced release of alkaline earth ions from the pyroxene lattice (calcium, magnesium, and strontium) resulting in an increase in pH, making the surrounding environment unsuitable for bacterial growth. The further addition of copper in co-substituted materials (Sr/Cu-DIO) significantly increased the antibacterial activity of Sr-DIO, which is somewhat expected as copper is known to produce reactive oxygen species that disrupt the membrane structure of E. coli, resulting in an antibacterial effect [60].

Biocompatibility and antibacterial activity are two conflicting aspects of bioactive ceramics and balancing these is a complex challenge in tissue engineering. Here, the Sr/Cu-DIO square strut scaffold we have prepared exhibits improved mechanical properties over the unmodified DIO scaffold and a good balance between biocompatibility and antibacterial properties. Based on these results, it can be seen that the Sr/Cu-DIO system offers promising scope towards tailoring modes of bioactivity in bone tissue engineering scaffolds and the ability to produce this material in the form of architected scaffolds offers further scope for the engineering of high-performance implants.

# 5. Conclusion

Strontium- and strontium/copper-doped diopside square strut bone tissue engineering scaffolds were successfully fabricated by robocasting. The results show that strontium and copper were readily incorporated in the diopside lattice, producing a decreased grain size and a more densified microstructure. Intrinsic mechanical properties of Vickers hardness and fracture toughness were improved in these materials by this substitution, exceeding the values of cancellous bone. Scaffolds produced with square cross-sectional strut profiles combine high porosity (68 - 71%), and compressive strengths, as the result of reduced stress concentration. Materials exhibited good apatite-forming ability, with cation substitution enhancing adhesion and proliferation of Saos-2 cells as well as antibacterial activity against E. coli. This study demonstrates how co-substitution and structural design can be implemented to customize and enhance the performance of bioactive diopside bone-tissue engineering scaffolds achieving unprecedented combinations of mechanical and multi-mode bioactive performance. The selection of strontium and copper has been validated here as an effective approach, and further co-substitutions, where multiple cations substitute at each of the two octahedral diopside sites may offer yet further avenues for materials design to explore in this system.

# Declaration of competing interest

The authors declare no conflict of interest.

# Acknowledgment

Shumin Pang greatly acknowledges the support from the China Scholarship Council (CSC, 201906780023). We thank Maria Unterweger at Technische Universität Berlin for carrying out the XPS measurement. We are also grateful to Viola Röhrs at Technische Universität Berlin for assisting us with the *in vitro* biological studies.